# Design of Intelligent Agents Based System for Commodity Market Simulation with JADE

R.Refianti [1], A.B. Mutiara[2], H. Gunawan[3]
*Faculty of Computer Science and Information Technology,
Gunadarma University, Indonesia*
E-mail: [1,2]{rina,amutiara}@staff.gunadarma.ac.id, [3] hendrabiter@gmail.com

## Abstract

A market of potato commodity for industry scale usage is engaging several types of actors. They are farmers, middlemen, and industries. A multi-agent system has been built to simulate these actors into agent entities, based on manually given parameters within a simulation scenario file. Each type of agents has its own fuzzy logic representing actual actors' knowledge, to be used to interpreting values and take appropriated decision of it while on simulation. The system will simulate market activities with programmed behaviors then produce the results as spreadsheet and chart graph files. These results consist of each agent's yearly finance and commodity data. The system will also predict each of next value from these outputs.

**Keywords:** Agent, JADE, Java, Fuzzy logic, Back propagation, Potato

## 1. Introduction

In potato commodity market, industries and farmers, as end buyer and raw producer respectively, hold big role in market activities that can give impact to each other. Industries need the farmers to fulfill their raw commodity requirement for continuous production. For industries, their production is vital activities to gain profit. At other side, farmers need to keep producing and sell their harvest revenue to have income. However, both farmers and industries are not the only one "seller" and "buyer" in market. Middlemen hold those both two roles, positioning themselves a competitor to farmers and industries and gain profit from it.

In market competition, each actor's assets, knowledge, and behaviors can be different to others thus giving different measures on the same situation. A simulation system will help by simulating the market activities and the produce the data to see if their current conditions can give positive impact in achieving their goal.

Agent Oriented Programming (AOP) is one of the best approaches to declare actual actors as system entities, called agent, and simulate their actions. AOP offers several advantages like message based communication, multi-behaviors support, life cycle management, and more [2]. The capability of agents can be improved by posing artificial intelligent backup to help them sense conditions of environment where they live and take appropriated actions regarding to it. Thus this type of agents are called intelligent agent [1]. Users have to supply actors' knowledge and scenario to the system. The system will be limited to give output data as it is, without any conclusion.

## 2. Methodology

The methodology that is used consists four phases in order [2]. Each phase has multiple steps inside them to be done in order too. The flow, phases, and steps of the methodology are illustrated in Fig. 1.

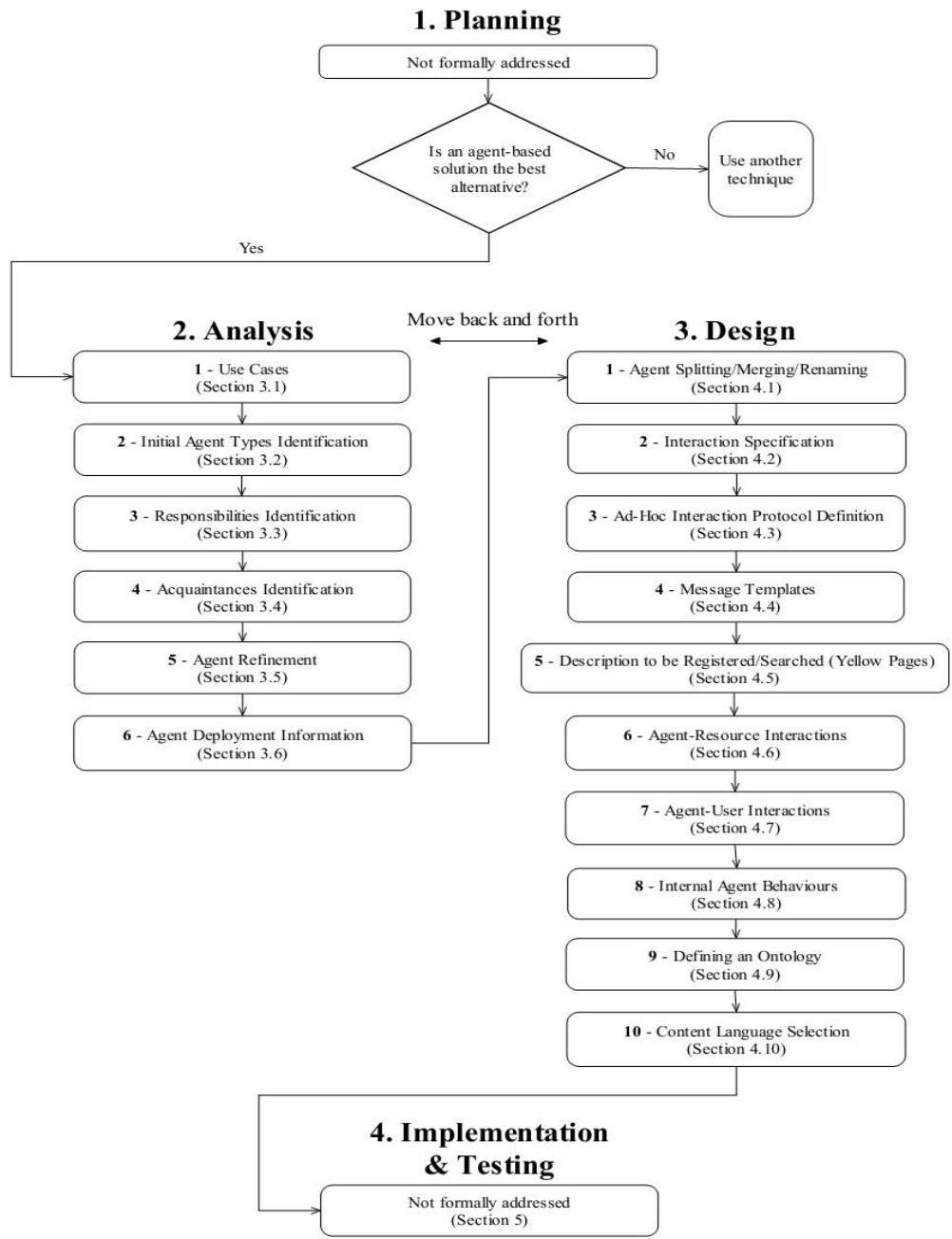

Figure 1. Methodology diagram [2]

This methodology is used corresponding to JADE (Java Agent Development) library which is used to built system's architecture. JADE is Java based library with ACL (Agent Communication Language) that is defined by FIPA (Foundation for Intelligent Physical Agent). In general, this ACL declares the messages performative and service type "yellow paging" as basic way for agents to communicate.

The methodology serves as a guide for the system designer when developing a system. In general, a software development methodology may comprise of [2]:
- A process, i.e. a sequence of phases and steps that guide the developer in building the system.
- A set of heuristic rules that support the developer in making relevant choices.
- A number of artifacts, i.e. diagrams, schemas or documents representing in graphical or textual form one or more models of the system.
- A suitable notation to be used in the artifacts.

- A set of patterns that can be applied to solve common situations.
- One or more tools that: automate, as much as possible, the phases and step specified in the process; force consistency between the models produced; highlight problems arising from incorrect design choices, when possible; generate code and documentation, etc.

The focus of the methodology is on the process and the artifacts that are produced. The described process covers the analysis phase and the design phase and is shown in Fig.1. The analysis phase is general in nature and independent of the adopted platform. Conversely, the design phase specifically assumes JADE as the implementation platform and focuses directly on the classes and concepts provided by JADE. Observing Fig.1, it can be seen that there is no strict boundary between the analysis and design phases. Moreover, the methodology is of an iterative nature, thus allowing the designer to move back and forth between the analysis and design phases and the steps therein.

At the end of the design phase, the developer should be able to progress straight to the implementation, which is where the actual coding occurs. In addition, most of this phase can probably be carried out by means of a proper tool which automates the implementation process. The planning stage, like implementation and testing, is not formally addressed in the methodology. However, for the sake of the methodology, a question is included (see Fig.1), which initially asks if the designer has made a rational decision on whether to use an agent-based solution. If the answer is yes, the designer moves on the analysis, while if the answer is no, the designer should seek an alternative solution.

## 3. Results and Discussion

### 3.1. Planning Phase

The system uses intelligent agent and Fuzzy logic is chosen as knowledge implementation since the actual actors may interpret some values into natural language. The fuzzy rules are written as FCL (Fuzzy Control Language) script and applied with help of jFuzzyLogic library and its built-in fuzzy inference system. The required input is simulation scenario describing some parameters including FCL scripts' location in form of spreadsheet file. At start, system will ask user to select this file in order to run. Finally, each agent will produce their finance and commodity data at the end of simulation, both in form of spreadsheet files (tabular) and picture files (charts) by using jExcel and jFreeChart libraries.

### 3.2. Analysis Phase

The following Use case diagrams are showing relation between farmer-middlemen, farmer-industries, and middlemen-industries respectively.

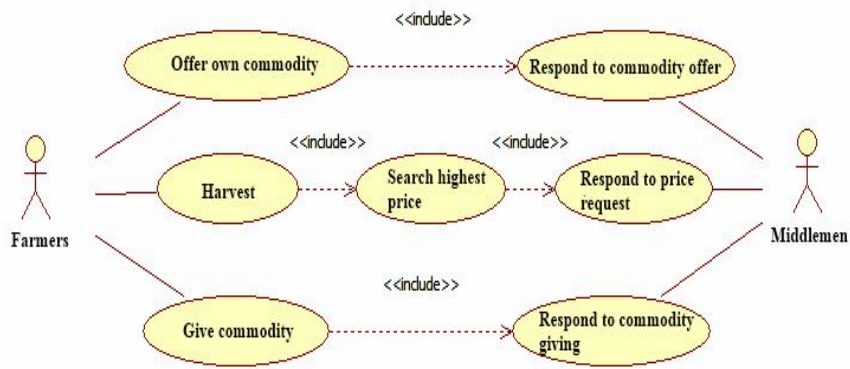

Figure 2. Use Case Diagram between Farmers and Middlemen

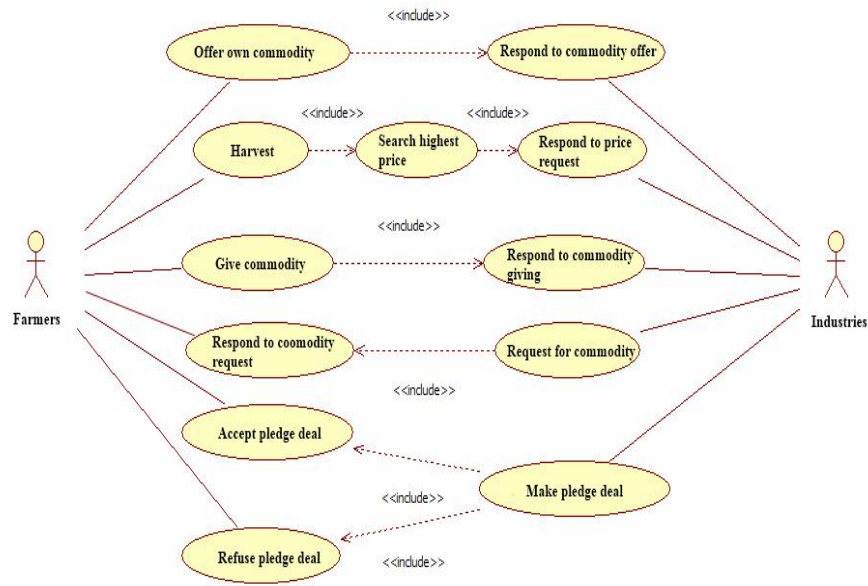

Figure 3. Use Case Diagram between Farmers and Industries

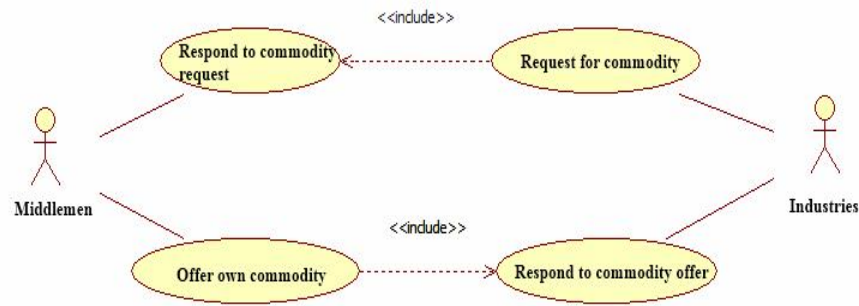

Figure 4. Use Case Diagram between Middlemen and Industries

Actual actors are represented by three kind of agents based on service they provide. Farmers, middlemen, and industries are "ProdusenAgent", "DistributorAgent", and "KonsumenAgent" agents respectively. Since the system needs to access the external resource scenario and FCL files, a transducer agent that not representing actual actors "MainAgent" is created. Fig. 2 shows relation between these agents kind.

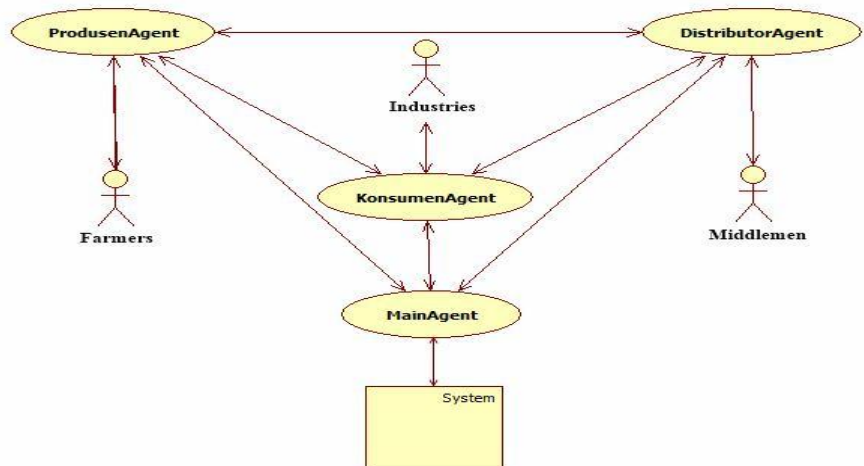

Figure 5. Agents relation

Table 1 below shows those agents' tasks.

**Table 1:** Tasks of agents

| Agent Type | Tasks |
|---|---|
| *ProdusenAgent* | Plant seed to field |
| | Harvest |
| | Respond to commodity request |
| | Respond to pledge deal |
| | Offer own commodity |
| | Search highest price for commodity |
| | Give harvest revenue to pledge dealer |
| *DistributorAgent* | Sort commodity according to its standard |
| | Respond to commodity offers |
| | Respond to price request |
| | Respond to commodity giving |
| | Respond to commodity request |
| | Offer own commodity |
| *KonsumenAgent* | Perform monthly production |
| | Sort commodity according to its standard |
| | Respond to commodity offers |
| | Respond to price request |
| | Respond to commodity giving |
| | Make pledge deal with *ProdusenAgent* |
| | Request for commodity |
| *MainAgent* | Load scenario file |
| | Parse scenario's parameters |
| | Do the dating count |
| | Synchronize date to all active agents |

### 3.3. Design Phase

This phase offers chance to improve agents relation and interaction protocol in case FIPA doesn't cover the need. After specify agents relation, a interaction specification table including messages template is made as shown in Table 2. All capitalized word refers to messages performative.

**Table 2:** Interaction specification table

| Agent | Interaction | IP | Role | With | When | Template |
|---|---|---|---|---|---|---|
| *ProdusenAgent* | Respond to commodity request | Offer | R | D, K | Get REQUEST message | QUERY_IF |
| | Respond to pledge deal | Seed | R | K | Get PROPOSE message | ACCEPT_PROPOSAL, REJECT_PROPOSAL |
| | Offer own commodity | Offer | I | D, K | Every month change and has stock | INFORM, INFORM_IF, AGREE, REFUSE |
| | Search highest price for | Pledge | I | D, K | Harvest time | CFP |

| | commodity | | | | | |
|---|---|---|---|---|---|---|
| | Give harvest revenue to pledge dealer | Pledge | I | D, K | Deal on given price | CONFIRM |
| *DistributorAgent* | Respond to commodity offers | Offer | R | P | Get INFORM message | INFORM_IF |
| | Respond to price request | Pledge | R | P | Get CFP message | REQUEST_WHEN |
| | Respond to commodity giving | Pledge | R | P | Get CONFIRM message | DISCONFIRM |
| | Respond to commodity request | Offer | R | K | Get REQUEST message | QUERY_IF |
| | Offer own commodity | Offer | I | K | Every month change and has stock | INFORM, INFORM_IF, AGREE, REFUSE |
| *KonsumenAgent* | Respond to commodity offer | Offer | R | P, D | Get INFORM message | INFORM_IF |
| | Respond to price request | Pledge | R | P | Get CFP message | REQUEST_WHEN |
| | Respond to commodity giving | Pledge | R | P | Get CONFIRM message | DISCONFIRM |
| | Make pledge deal with *ProdusenAgent* | Seed | I | P | Every month | PROPOSE, AGREE, REFUSE |
| | Request for commodity | Offer | I | P, D | Every month when stock isn't enough to perform production | REQUEST, AGREE, REFUSE |
| *MainAgent* | Synchronize date to all active agents | Detailed Year | I | P, D, K | Every month | SUBSCRIBE |

To interact with user, the system has its own graphical user interface (GUI) showing finance and commodity charts. The GUI also shows some agent's actions log for debugging case and agent's current status as described in stock, money, and field values. User gives the scenario input through "*MainAgent*". The scenario file has two sections of parameters. The first section as described on Table 3 is used to determine how the simulation will work globally. A scenario file only has one of these sections.

**Table 3**:   Global parameters

| Title | Description |
|---|---|
| Specification block counts | Number of how many specification blocks that need to load |
| Start year | Year of start |
| Simulation duration (year) | How long the simulation |
| Harvest ratio (Revenue/seed) | Number of harvest revenue (Kg) produced by 1 Kg seed |
| Kg seed per Ha | Number of seed (Kg) planted in 1 Ha field |
| Autonom (boolean) | Specify whether simulation will run automatically (1) or manually (0) |

The second section is used to give parameters to agent. This section can appear multiple times in one scenario file. The number of this section must be specified at the start of scenario, in other words first global parameter. The section is described in Table 4.

**Table 4**: Agents parameters

| Name | Description |
|---|---|
| GUI | Specify whether agent(s) has GUI (1) or not (0) |
| Name | Name for agent(s) |
| Role | Service type of agent(s): "produsen" (0), "distributor" (2), "konsumen" (3) |
| Stock at start (kg) | Number of stock at start |
| Money at start (Rp) | Number of money at start |
| Seed at start (Kg) | Number of seed at start |
| Minimum diameter (cm) | Lower value of diameter to pass selection |
| Maximum diameter (cm) | Upper value of diameter to pass selection |
| Production usage (kg) | Number of commodity needed to perform one time production |
| Production income (Rp/kg) | Price for 1 Kg of processed commodity |
| Market income (Rp/kg) | Price for 1 Kg of commodity when sold in traditional market |
| Normal buy price (Rp/Kg) | Starting price when buying |
| Normal pledge price (Rp/Kg) | Starting price when pledging |
| Harvest failure chance (%) | Probability for harvest revenue number to randomly goes down |
| Field (Ha) | Owned field |
| Plant cost (Rp/Ha) | Cost to plant in 1 Ha field |
| FCL path | File path relatively to system's location for FCL script |
| Number of agents | Number of agent(s) using this specification |

Each agent has some instance of JADE's Behavior class to describe an action of actual actor into algorithm. These behaviors can be executed in parallel or sequence depends on the need. The lists of behaviors representing actual actors' actions are mentioned by Table 5.

**Table 5**: Agent's actor representing behaviors

| Behavior's Name | Represented Action | Type of agent |
|---|---|---|
| InformingBev | Offer own commodity | ProdusenAgent, DistributorAgent |
| InformedBev | Respond to commodity offer | DistributorAgent, KonsumenAgent |
| QueryingBev | Request for commodity | KonsumenAgent |
| QueriedBev | Respond to commodity request | ProdusenAgent, DistributorAgent |
| SellBev | Offer own commodity to respond request | ProdusenAgent, DistributorAgent |
| PledgingBev | Make pledge deal | KonsumenAgent |
| PledgedBev | Respond to pledge deal | ProdusenAgent |
| GivingBev | Give commodity | ProdusenAgent |
| GivedBev | Respond to commodity giving | DistributorAgent, KonsumenAgent |
| DivergingBev | Search highest price | ProdusenAgent |
| DivergedBev | Respond to price request | DistributorAgent, KonsumenAgent |

Beside those behaviors, there are some which aren't representing actual actors' actions. These behaviors are related to agents' activities like calculate date and pool received message. They are listed in Table 6.

**Table 6**: Agent's system related behaviors

| Behavior's Name | Purpose | Type of agent |
|---|---|---|
| CyclicBehavior | Used to pool received message and check its performative then call appropriate behavior to handle it. | ProdusenAgent, DistributorAgent, KonsumenAgent |
| DelayBehavior | Used to update date by increasing month count after some time. It'll be done as long as simulation running. | MainAgent |

For ontology, the system will use object serialization as its protocol. To do this, some classes which will be content language like Price, Offer, and Seed have to implement Serializable interface. Instance of these classes will be set to ACLMessage before sent. ACLMessage itself is JADE's class that support performative and object use as content rather plain string.

### 3.4. Implementation and Testing Phase

A test is performed by using a dummy scenario and three kinds of knowledge bases, one for each type of agents. The scenario specifies six block specification for six agents (one block per agent) and put them in group of two. Therefore each type of agents has two agents instance and shares same knowledge base. The complete list of scenario can be seen in Table 6. This table uses scenario's dummy values for test. Note that this table doesn't represent how to write them on the actual scenario file because there's difference in format. In order to optimize, several operation parameters for Java Virtual Machine (JVM) and JADE class loader are given as listed in Table 7.

**Table 6**: Complete list of scenario's dummy

| Parameters | Values | |
|---|---|---|
| Specification block counts | 6 | |
| Start year | 2002 | |
| Simulation duration (year) | 8 | |
| Harvest ratio (Revenue/seed) | 2 | |
| Kg seed per Ha | 1000 | |
| Autonom (boolean) | 1 | |
| | | |
| Group | Produsen | |
| GUI | 1 | 1 |
| Name | P1 | P2 |
| Role | 0 | 0 |
| Stock at start (kg) | 10 | 5 |
| Money at start (Rp) | 35000 | 100000 |
| Seed at start (Kg) | 20 | 34 |
| Minimum diameter (cm) | 0 | 0 |
| Maximum diameter (cm) | 0 | 0 |
| Production usage (kg) | 0 | 0 |
| Production income (Rp/kg) | 0 | 0 |
| Market income (Rp/kg) | 1000 | 1600 |
| Normal buy price (Rp/Kg) | 5300 | 5100 |
| Normal pledge price (Rp/Kg) | 0 | 0 |
| Harvest failure chance (%) | 0.45 | 0.57 |

| Field (Ha) | 2.3 | 5.3 |
|---|---|---|
| Plant cost (Rp/Ha) | 50000 | 56400 |
| FCL path | script/produsen.fcl ||
| Number of agents | 1 | 1 |
| **Group** | **Distributor** ||
| GUI | 1 | 1 |
| Name | D3 | D4 |
| Role | 1 | 1 |
| Stock at start (kg) | 12 | 3 |
| Money at start (Rp) | 250000 | 37600 |
| Seed at start (Kg) | 0 | 0 |
| Minimum diameter (cm) | 4.9 | 3.9 |
| Maximum diameter (cm) | 6.3 | 5.7 |
| Production usage (kg) | 0 | 0 |
| Production income (Rp/kg) | 0 | 0 |
| Market income (Rp/kg) | 1200 | 1200 |
| Normal buy price (Rp/Kg) | 5700 | 5500 |
| Normal pledge price (Rp/Kg) | 0 | 0 |
| Harvest failure chance (%) | 0 | 0 |
| Field (Ha) | 0 | 0 |
| Plant cost (Rp/Ha) | 0 | 0 |
| FCL path | script/distributor.fcl ||
| Number of agents | 1 | 1 |
| **Group** | **Konsumen** ||
| GUI | 1 | 1 |
| Name | K5 | K6 |
| Role | 2 | 2 |
| Stock at start (kg) | 500 | 830 |
| Money at start (Rp) | 400000 | 100000 |
| Seed at start (Kg) | 80 | 220 |
| Minimum diameter (cm) | 5.1 | 6.5 |
| Maximum diameter (cm) | 6 | 8.79 |
| Production usage (kg) | 150 | 125 |
| Production income (Rp/kg) | 3000 | 4100 |
| Market income (Rp/kg) | 1500 | 900 |
| Normal buy price (Rp/Kg) | 6000 | 6400 |
| Normal pledge price (Rp/Kg) | 6200 | 7500 |
| Harvest failure chance (%) | 0 | 0 |
| Field (Ha) | 0 | 0 |
| Plant cost (Rp/Ha) | 0 | 0 |
| FCL path | script/konsumen.fcl ||

| Number of agents | 1 | 1 |

**Table 7**: Operation parameters

| Parameter | Description |
|---|---|
| -Xms512m | Allocate 512MB of RAM to JVM |
| -Xmx1024m | Allocate maximal 1024MB of RAM to JVM for additional usage |
| jade_core_messaging_MessageManager _maxqueuesize 50000000 | Tell JADE to allocate 50MB of RAM as message queue |

The results of simulation for two agents of "ProdusenAgent" are shown in below figures. Fig.6 shows complete GUI of the agents and commodity charts at same time. The finance charts hidden inside the scroll area can be seen in Fig.7.

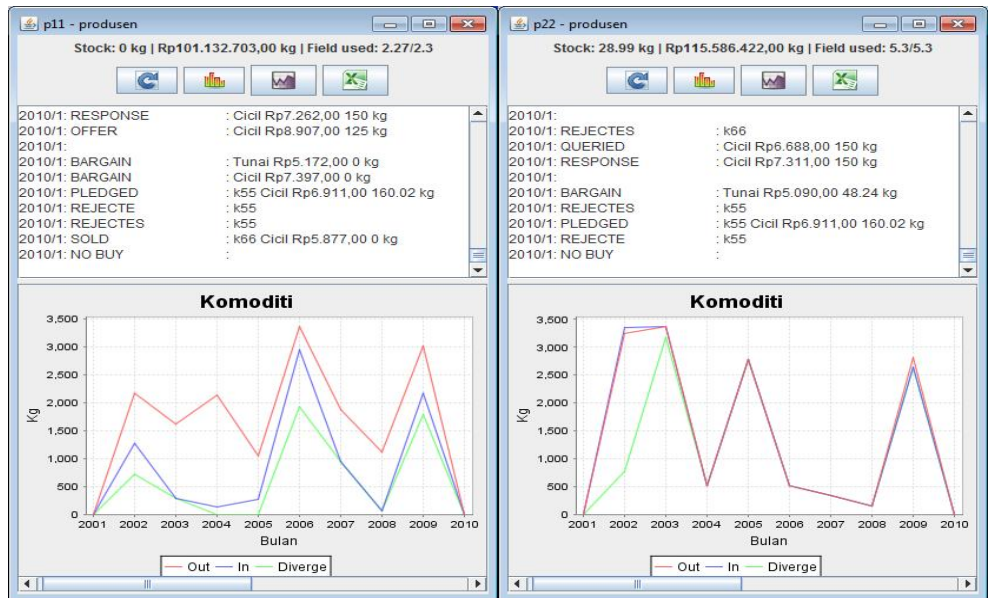

Figure 6. Commodity charts of "ProdusenAgent" agents

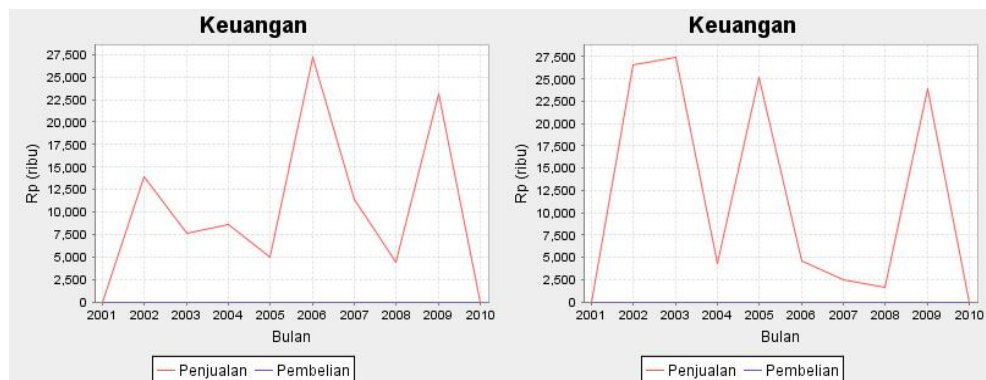

Figure 7. Finance charts of "ProdusenAgent" agents

The results of simulation for two agents of "DistributorAgent" are shown both in commodity (Fig. 8) and in finance (Fig. 9) charts too.

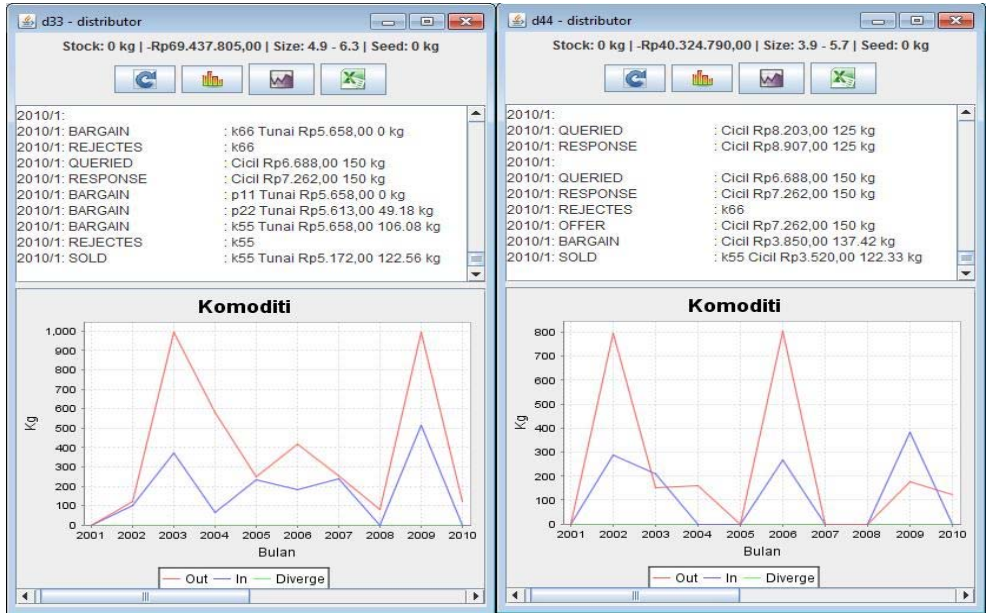
Figure 8. Commodity charts of "DistributorAgent" agents

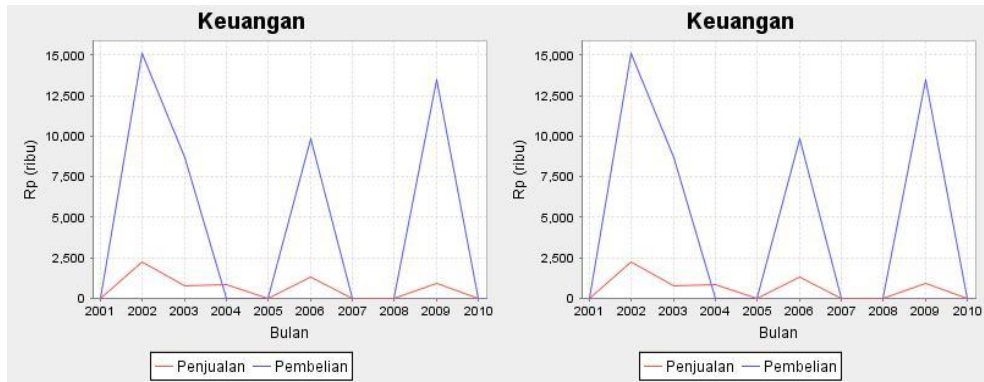
Figure 9. Finance charts of "DistributorAgent" agents

Finally, the output of "KonsumenAgent" agents are shown by Fig.10 and Fig.11.

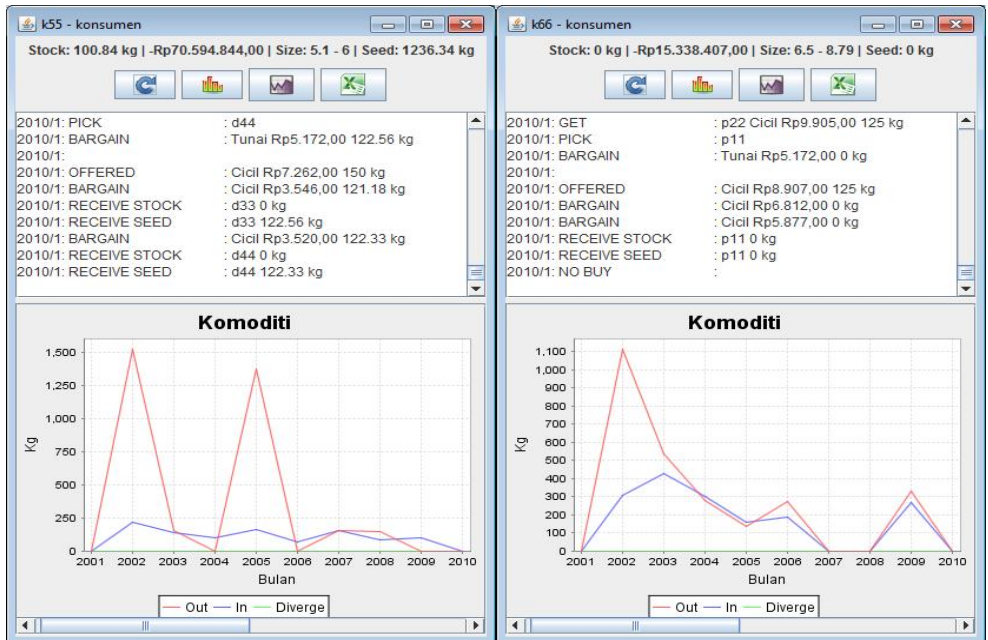
Figure 10. Commodity charts of "KonsumenAgent" agents

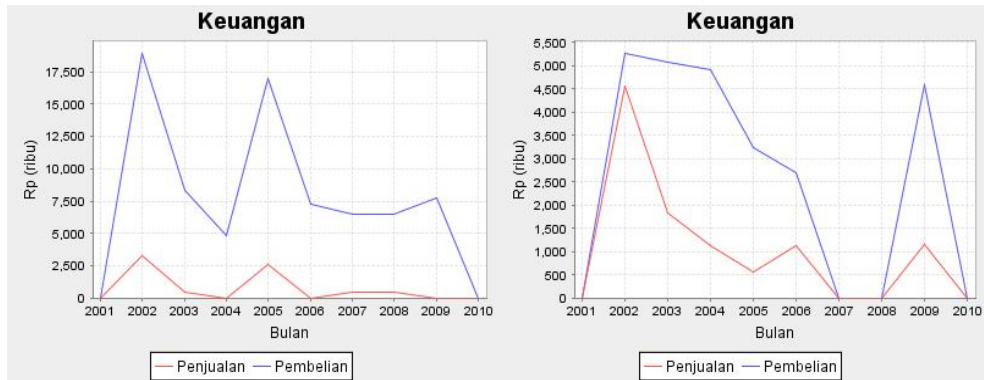
Figure 11. Finance charts of "KonsumenAgent" agents

For additional test, the system is also strained by simulating some numbers of agents instances (Table 8), including previous dummy test as test no. I. This is conducted to see how well the system's performance is. Table 9 shows specification of the testing machine where all test performed. Memory and processor (CPU) usages that are used during the test are listed by Table 10.

**Table 8**: Number of agent's instances for strain test

| Test No. | Number of "ProdusenAgent" instance | Number of "DistributorAgent" instance | Number of "KonsumenAgent" instance | Total |
|---|---|---|---|---|
| I | 2 | 2 | 2 | 6 |
| II | 7 | 7 | 7 | 21 |
| III | 12 | 12 | 12 | 36 |
| IV | 17 | 17 | 17 | 51 |
| V | 22 | 22 | 22 | 66 |

**Table 9**: Testing machine

| Parameter | Description |
|---|---|
| Operating System | Windows 7 Ultimate 32 bit |
| CPU | AMD E-450 2 CPU @1.6GHz |
| RAM | 2048MB, 384MB Shared |
| Storage | ST950032 SATA |
| JDK/JRE version | Java 6 update 29 |

**Table 10**: Strain test results

| Test No. | Memory usage | | CPU usage | Duration |
|---|---|---|---|---|
| | Min (Byte) | Max (Byte) | | |
| I | 21.635.680 | 156.129.920 | 24,90% | 5 m 46 s |
| II | 19.316.896 | 159.294.232 | 98,30% | 5 m 59 s |
| III | 17.112.774 | 164.188.432 | 100,00% | 5 m59 s |
| IV | 24.967.448 | 591.403.360 | 100,00% | 6 m 44 s |
| V | 49.914.848 | 1.360.807.024 | 100,00% | 19 m 54 s |

## 4. Conclusions

The built system can run simulation based on given scenario then produce output each agent's yearly finance and commodity data in form of spreadsheet and chart graph files. The system is also able to add additional value for spreadsheet files as prediction. The output is limited to raw data without any further analysis or conclusions, which are expected to come from appropriate market experts. Inheriting advantage of agent based application, system extension efforts like adding new agents or new behaviors is possible. For further research, some steps can be taken like implementing self adapting logic, performance optimization, and improvement for better quality of results.

## References

[1] Fabio Bellifemine., Giovanni Caire., Dominic Greenwood., 2007, *Developing multi-agent systems with JADE*, John Wiley and Sons, Ltd, West Sussex, England.

[2] Magid Nikraz., Giovanni Caire., Parisa A. Bahri., 2xxx, *A methodology for the Analysis and Design of Multi-Agent Systems using JADE*, Telecom Italia Lab, Via Reiss Romoli, Turin, Italy 10148.